\documentclass[12pt,preprint]{aastex62}
\usepackage{color}
\usepackage{graphicx}
\usepackage{longtable}
\usepackage{multirow}
\usepackage{url}
\usepackage{subfigure}
\usepackage{amsmath}
\usepackage{lipsum}

\begin{document}

%\title{Searching for the missing baryons in the intergalactic medium with localized fast radio bursts}
\title{Finding the missing baryons in the intergalactic medium with localized fast radio bursts}

\author[0000-0001-9445-9215]{K. B. Yang}
\affiliation{School of Astronomy and Space Science, Nanjing University, Nanjing 210093, China}

\author[0000-0001-6021-5933]{Q. Wu}
\affiliation{School of Astronomy and Space Science, Nanjing University, Nanjing 210093, China}

\author[0000-0003-4157-7714]{F. Y. Wang}
\affiliation{School of Astronomy and Space Science, Nanjing
	University, Nanjing 210093, China} \affiliation{Key Laboratory of
	Modern Astronomy and Astrophysics (Nanjing University), Ministry of
	Education, Nanjing 210093, China}

\correspondingauthor{F. Y. Wang}
\email{fayinwang@nju.edu.cn}
	
\begin{abstract}
The missing baryon problem is one of the major unsolved problems in astronomy. Fast radio bursts (FRBs) are bright millisecond pulses with unknown origins. The dispersion measure of FRBs is defined as the electron column density along the line of sight, and accounts for every ionized baryon. Here we measure the baryon content of the Universe using 22 localized FRBs. Unlike previous works that fixed the value of dispersion measure of FRB host galaxies and ignored the inhomogeneities of the intergalactic medium (IGM), we use the probability distributions of dispersion measures contributed by host galaxies and IGM from the state-of-the-art IllustrisTNG simulations. We derive the cosmic baryon density of $\Omega_b=0.0490^{+0.0036}_{-0.0033}$ (1$\sigma$), with a precision of 7.0\%. This value is dramatically consistent with other measurements, such as the cosmic microwave background and Big Bang nucleosynthesis. Our work supports that the baryons are not missing, but residing in the IGM.
\end{abstract}


\section{Introduction}

The observation of cosmic microwave background (CMB) by Planck Collaboration shows that more than 95\% of our universe is composed of dark energy and dark matter, while the rest part of less than 5\% is baryonic matter \citep{2020A&A...641A...6P}. In the late-time Universe, approximately $17\%$  of the expected baryons are observed to reside in the collapsed phase \citep{Fukugita1998}, including galaxies, groups, and clusters. Numerical simulations \citep{Cen1999} and observations \citep{Fukugita2004} are performed to study the baryon distribution. However, the fraction of baryons observed in the photoionized Ly$\alpha$ forest and warm-hot intergalactic medium (WHIM) traced by highly ionized states oxygen is only about 53\%, which left a vacancy of about 30\% \citep{shull12}. In recent years, some baryons are found in various ways. \cite{2019A&A...624A..48D} found additional $11 \pm 7\%$ baryons with the Sunyaev-Zel’dovich effect from filaments, and the still missing fraction became $18 \pm 16\%$. Some missing baryons had been found in the warm–hot intergalactic medium \citep{2018Natur.558..406N}.
%However, this problem has not been solved.
Most missing baryons may be residing in diffuse IGM, which is too faint to detect \citep{McQuinn2016}. Finding the missing baryon is important for us to understand many crucial problems, such as galaxy evolution.

Fast radio bursts (FRBs) are bright millisecond pulses \citep{2019A&ARv..27....4P, 2019ARA&A..57..417C,xiao21}. Since the first burst, FRB 010724 was discovered in 2007 \citep{2007Sci...318..777L}, over 600 FRBs have been detected \citep{chime21}. When FRBs travel through cold plasma, there will be a time difference between their components of different frequencies, which can be used to define the dispersion measure (DM). DM is defined as the integral of free electron number density along the line of sight. Owing to their large DM in excess of the Galactic value, they are suggested to be extragalactic or cosmological. It is estimated that the majority of DM contribution comes from IGM. The IGM is comprised of highly ionized plasma, which means that the baryon density can be derived from the electron density.  %After a few simple assumptions, the relation between the DM of an FRB contributed by IGM and its redshift can be derived.
Based on the average DM-$z$ relation, short timescales and cosmological origin make them a powerful probe to study the universe and fundamental physics, including measuring Hubble constant \citep{Li18, wq22, Hagstotz2022, James22} and Hubble parameter \citep{WU2020}, measuring cosmic proper distance \citep{Yu17}, measuring dark energy \citep{Zhou14, Walters2018, qiu22}, bounding the photon rest mass \citep{2021PhLB..82036596W}, measuring reionization history \citep{Zheng14,zzj21}, testing the weak
equivalence principle \citep{Wei2015}, probing compact dark matter \citep{2016PhRvL.117i1301M,Wang2018}, and finding ``missing" baryons \citep{McQuinn14,Walters2018,Walters2019,Macquart20,Li20, Dai2021,WangB2022}.
\cite{Macquart20} performed the first observational constraint on the cosmic baryon density $\Omega_b$ using five FRBs, and found $\Omega_b=0.051^{+0.021}_{-0.025}$ at 95\% confidence level, which is consistent with the value derived from the CMB and from Big Bang nucleosynthesis (BBN).

However, there are some obstacles in practice with cosmological applications of FRBs. First, the number of FRBs with measured redshifts is too small, which can be improved with increasing localized FRBs. Thanks to the high event rate, i.e., $10^3-10^4$ per day all sky \citep{2013Sci...341...53T, Caleb16}, the increasing advanced telescopes will localize more FRBs. The second one is that we know little about $\rm DM_{host}$, the DM contribution from the host galaxy. It is also degenerated with $\rm DM_{IGM}$, the DM contribution from IGM. How to model $\rm DM_{host}$ is a crucial problem. Third, the IGM inhomogeneity causes the fluctuations of DM$_{\rm IGM}$ along different lines of sight. From cosmological simulations, the variance around the mean DM$_{\rm IGM}$ can be up to 70\% \citep{McQuinn14, Pol2019, zzj21}. In most previous works, the value of $\rm DM_{host}$ for all host galaxies is treated as a constant, and the fluctuations of $\rm DM_{IGM}$ are ignored. A more appropriate method to describe $\rm DM_{IGM}$ and $\rm DM_{host}$ is needed in the cosmological application of FRBs.

In this letter, we measure the cosmic baryon density with 22 localized FRBs through the DM-$z$ relation. Unlike previous works which assume the $\rm DM_{host}$ as a constant, we use the probability distributions of it from the state-of-the-art IllustrisTNG simulations \citep{zgq20}. The effect of different properties of host galaxies on $\rm DM_{host}$ is also taken into consideration. Meanwhile, the fluctuations of DM$_{\rm IGM}$ along different lines of sight are also taken into account \citep{zzj21}.

The letter is organized as follows. In section 2, we introduce the probability distributions of $\rm DM_{host}$ and $\rm DM_{IGM}$, and the method to derive $\Omega_b$. In section 3, we constrain the value of $\Omega_b$ through the Monte Carlo Markov Chain (MCMC) analysis. Discussion and conclusions are given in section 4.

\section{Method}\label{sec2}

%When an FRB is observed, its DM can be measured with high precision.
The observed DM of FRBs can be divided into four parts
\begin{equation}
 {\rm DM}_{\rm obs} ={\rm DM}_{\rm MW} +{\rm DM}_{\rm halo} +{\rm DM}_{\rm IGM} + \frac{{\rm DM}_{\rm host}}{1+z},
\label{DMobs}
\end{equation}
where $\rm DM_{MW}$ is contributed by the interstellar medium of the Milky Way, $\rm DM_{halo}$ is contributed by the Milky Way halo, and $\rm DM_{IGM}$ refers to the contribution of IGM, and $\rm DM_{host}$ is contributed by the host galaxy. For the first part $\rm DM_{MW}$, some models have been built, including NE2001 and YMW16 models \citep{NE2001,YMW16}. In this work, we use the NE2001 model to estimate the value of $\rm DM_{MW}$.
After deducing the $\rm DM_{MW}$ from $\rm DM_{obs}$, we can get $\rm DM^{\prime}_{FRB}$ = $\rm DM_{IGM}$ + $\rm DM_{host}$ + $\rm DM_{halo}$. Only the part $\rm DM_{IGM}$ contains the cosmological information, including the cosmic baryon density $\Omega_b$. However, it's difficult to separate it from $\rm DM_{host}$ and $\rm DM_{halo}$, which is the key point for the cosmological applications of FRBs. Besides, due to the matter inhomogeneity in the IGM and the diverse properties of host galaxies, both of them can not be accurately determined.

In a flat $\Lambda$CDM universe, the average $\rm DM_{IGM}$ is \citep{Deng2014,Macquart20}
\begin{equation}\label{dmigm}
\langle {\rm DM_{\rm IGM}}\rangle = \frac{3c\Omega_b H_0}{8\pi G m_p }\int_{0}^{z_{ FRB}}\frac{f_{\rm  IGM}(z)f_e(z)(1+z)}{\sqrt{\Omega_m(1+z)^3+ \Omega_\Lambda}}dz,
\end{equation}
where $m_p$ is the rest proton mass, $\Omega_m$ is the cosmic matter density, and the electron fraction is $f_e(z) = Y_H X_{e, H}(z) + \frac{1}{2} Y_{He}X_{e, He}(z) = 7/8$. For a flat universe, $\Omega_m=1-\Omega_\Lambda$. $f_{\rm IGM}(z)$ is the fraction of baryons in the IGM. At present, we know little about the evolution model of $ f_{\rm IGM}(z)$ with redshift $z$. $ f_{\rm IGM}(z)$ could increase with redshifts as massive halos grow abundant with the universe evolving. It's estimated that the value of $ f_{\rm IGM}$ is 0.82 at low redshifts and 0.9 at $z > 1.5$ \citep{2009RvMP...81.1405M,shull12}. For the Hubble constant $H_0$, a fiducial value of 70 km/s/Mpc is used. %Therefore, in our work, we try to estimate $ f_{IGM}(z)$ in two different models: $f_{\rm IGM}$ is a constant, the other it slowly increases with redshift. Here we follow \cite{Li19} and parameterize the slow growth with the function: $ f_{IGM}(z) = f_{IGM,0}(1+\alpha z/(1+z))$, in which $\alpha$ is a positive constant and $ f_{\rm IGM,0}(z)$ is the fraction of baryons at $z = 0$.

\begin{figure}
\centering
\includegraphics[width=1.0\textwidth]{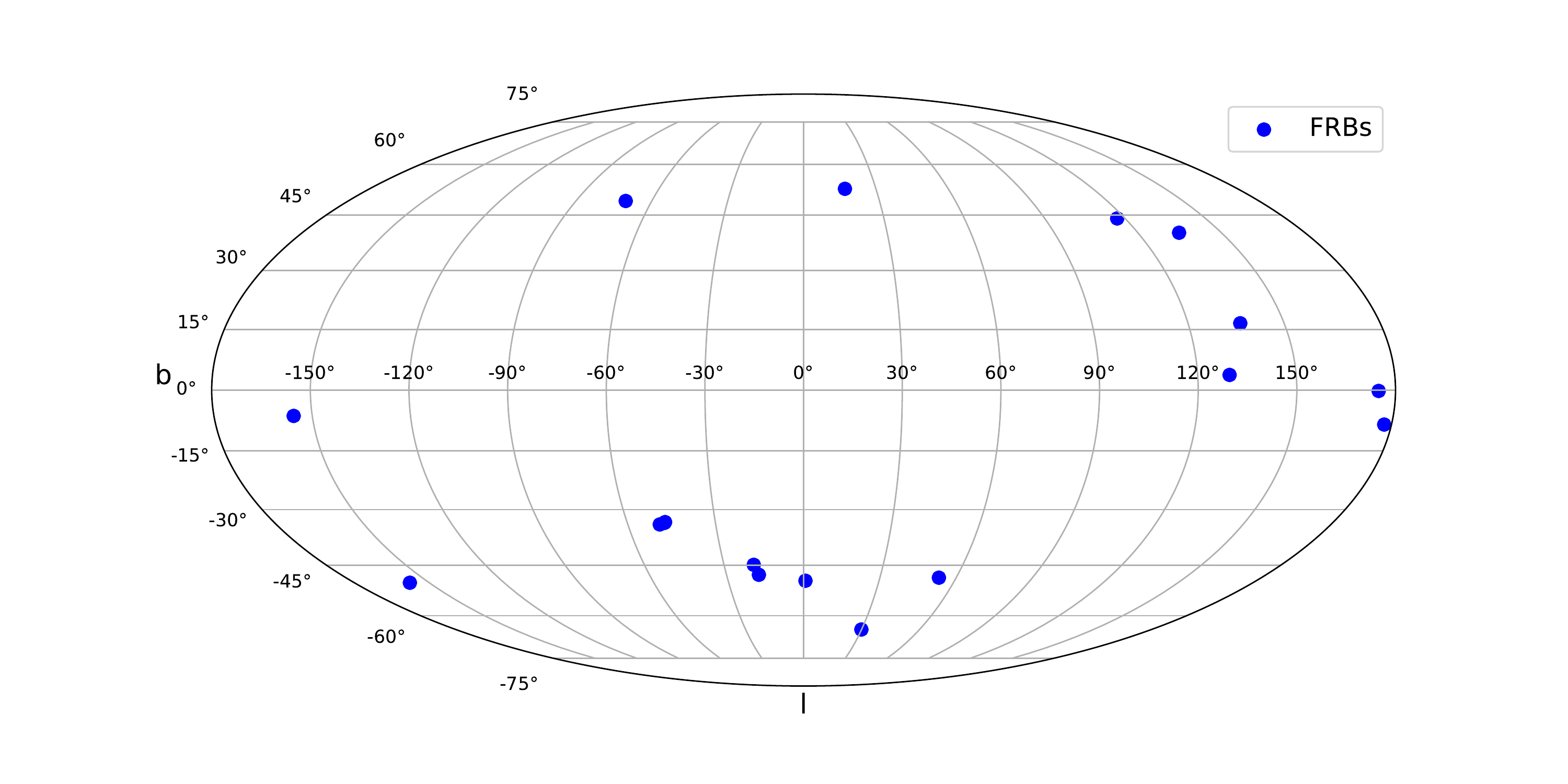}
\caption{\label{mollweide} The coordinates of 18 localized FRBs in the galactic coordinate system. There is no FRB with both low galactic latitudes and low galactic longitudes.}
\end{figure}

As mentioned above, it is difficult to determine the exact value of $\rm DM_{halo}, DM_{host}$ and DM$_{\rm IGM}$. So we focus on the probability distributions of them.
The $\rm DM_{halo}$ contributed by the halo is estimated to be $ 50 \sim 80\rm \  pc \ cm^{-3}$ \citep{2019MNRAS.485..648P}.
According to this estimation, we assume that the distribution of $\rm DM_{halo}$ is a Gaussian distribution with the standard deviation $\sigma_{\rm halo}=15 ~\rm pc\ cm^{-3}$ and the mean value $\mu_{\rm halo}=65~ \rm pc\ cm^{-3}$ \citep{wq22}, i.e.,
\begin{equation}\label{dmhalo}
P_{\rm halo}(\rm DM_{\rm halo}| \mu_{halo}, \sigma_{halo}) = \frac{1}{\sigma_{\rm halo}\sqrt{2\pi}} \exp[-\frac{(DM_{\rm halo}-\mu_{\rm halo})^2}{2\sigma^2_{\rm halo}} ].
\end{equation}
It is worth noting that $\rm DM_{halo}$ is larger than $ 30\rm\  pc \ cm^{-3}$, and should be smaller than $ 100\rm\ pc \ cm^{-3}$ for FRBs at high galactic latitudes and longitudes \citep{yamasaki2020}. Positions of FRBs with known coordinates are shown in figure \ref{mollweide}. All of them satisfy this condition. For conservative estimations, the range of $\rm DM_{halo}$ is adopted as $ 30 \sim 100\rm \  pc \ cm^{-3}$ in our calculation \citep{yamasaki2020}.

For localized FRBs, we can calculate the average $\langle {\rm DM_{\rm IGM}}\rangle$ with equation (\ref{dmigm}). The probability distribution of DM$_{\rm IGM}$ can be described as a long-tailed quasi-Gaussian function of the average value \citep{McQuinn14, Macquart20,zzj21},
\begin{equation}\label{PIGM}
P_{\rm IGM}(\Delta) = A \Delta^{-\beta} \exp  [-\frac{(\Delta^{-\alpha}-C_0)^2}{2\alpha^2\sigma^2_{\rm DM}} ], \Delta > 0,
\end{equation}
where $\Delta =\rm DM_{IGM}/\langle DM_{IGM}\rangle$, $\alpha$ and $\beta$ are related to profile scales and their best fits are $\alpha = 3$ and $\beta = 3$ \citep{Macquart20}. $\sigma_{DM}$ is the effective deviation caused by the inhomogeneities of IGM. \cite{zzj21} have given the best fits of $C_0$, $A$ and $\sigma_{DM}$ at several different redshifts derived from the IllustrisTNG simulation. These redshifts do not include the redshifts of the FRBs used in this work.
Using the cubic spline interpolation method, we get $C_0$, $A$ and $\sigma_{DM}$ as a function of redshift. Then, the values of $C_0$, $A$ and $\sigma_{DM}$ at given redshifts are derived.

A log-normal function can well describe the distribution of DM$_{\rm host}$ \citep{Macquart20,zgq20},
\begin{equation}\label{Phost}
P_{\rm host}({\rm DM_{ host}}| \mu, \sigma) = \frac{1}{{\rm DM_{ host}} \sigma \sqrt{2\pi}} {{\exp}}[-\frac{({\rm ln} {\rm DM_{ host}}-\mu)^2}{2\sigma^2}],
\end{equation}
where $\mu$ and $\sigma$ are free parameters. For this distribution, the expected value and standard deviation are $e^\mu$ and $e^{2\mu+{\sigma}^2}[e^{\sigma^2}-1]$, respectively. \cite{zgq20} have given the best fitting values of $\mu$ and $\sigma$ through the state-of-the-art IllustrisTNG simulations. In their work, FRBs are divided into three types according to the observational properties of host galaxies: repeating FRBs like FRB 121102 (Type I), repeating FRBs like FRB 180916 (Type II) and non-repeating FRBs (Type III). Repeating bursts may originate from newborn neutron stars \citep{wang20}. FRB 121102 is localized to a dwarf galaxy with a low star formation rate, hence repeating FRBs like FRB121102 may be from dwarf galaxies with low metallicity \citep{Chatterjee2017}. The host galaxy of FRB 180916 is a spiral galaxy, which is different from that of FRB 121102. It was born in a star-forming region \citep{2020Natur.577..190M}. Therefore repeating FRBs like FRB 180916 are likely to reside in the star-forming region. Non-repeating FRBs may originate from the merger of compact binaries \citep{2013PASJ...65L..12T,Wang2016}, and their locations are far from the galaxy centers \citep{2019Sci...365..565B,2019Sci...366..231P}. %Because of these different properties, $\mu$ and $\sigma$ may be different.
In our calculation, as before, we use their results to get $\mu$ and $\sigma$ at given redshifts by cubic spline interpolation.

For an FRB with a given redshift $z$ and DM$^{\prime}_{\rm FRB,i}$, we can calculate the total probability density function (PDF) through
\begin{equation}\label{Ptotal}
\begin{split}
P_{i}\left({\rm DM}_{{\rm FRB}^{\prime}, i } \mid z_i\right) = & \int_{30}^{{100}} \int_{0}^{{\mathrm{DM}}^{\prime}_{ {\rm FRB},i}-\rm DM_{halo}}  P_{\text {IGM}}(\rm DM_{{IGM} }) P_{\text {halo}}({DM}_{halo}|\mu_{halo}, \sigma_{ halo})   \\
& P_{\text {host }}({\text {DM}^{\prime}_{{\rm FRB}, i }-{\rm DM}_{{\rm IGM} }-{\rm DM}_{\rm halo}}|\mu, \sigma_{ host})d{\rm DM_{\rm IGM}}d{\rm DM_{\rm halo}}.
\end{split}
\end{equation}
At last, we obtain the joint likelihood function by multiplying the PDF of each FRB
\begin{equation}\label{likelihood}
 \mathcal{L} = \prod\limits_{i=1}^{N_{\mathrm{FRB}}} P_{i}\left({\rm DM}^{\prime}_{{\rm FRB}, i } \mid z_i\right).
\end{equation}
%For the first case, i.e., $f_{IGM}$ is independent of redshift, the fitting of $f_{IGM}$ can be converted into a one-dimensional problem. For the second case, in which $f_{IGM}$ evolves with redshift, the Monte Carlo Markov Chain (MCMC) method is used to estimate $f_{\rm IGM}$ and the evolution parameter $\alpha$. %More details of our fitting will be introduced in the next section.
\section{Data and Results}\label{sec3}
\begin{figure}
\centering
\includegraphics[width=0.8\textwidth]{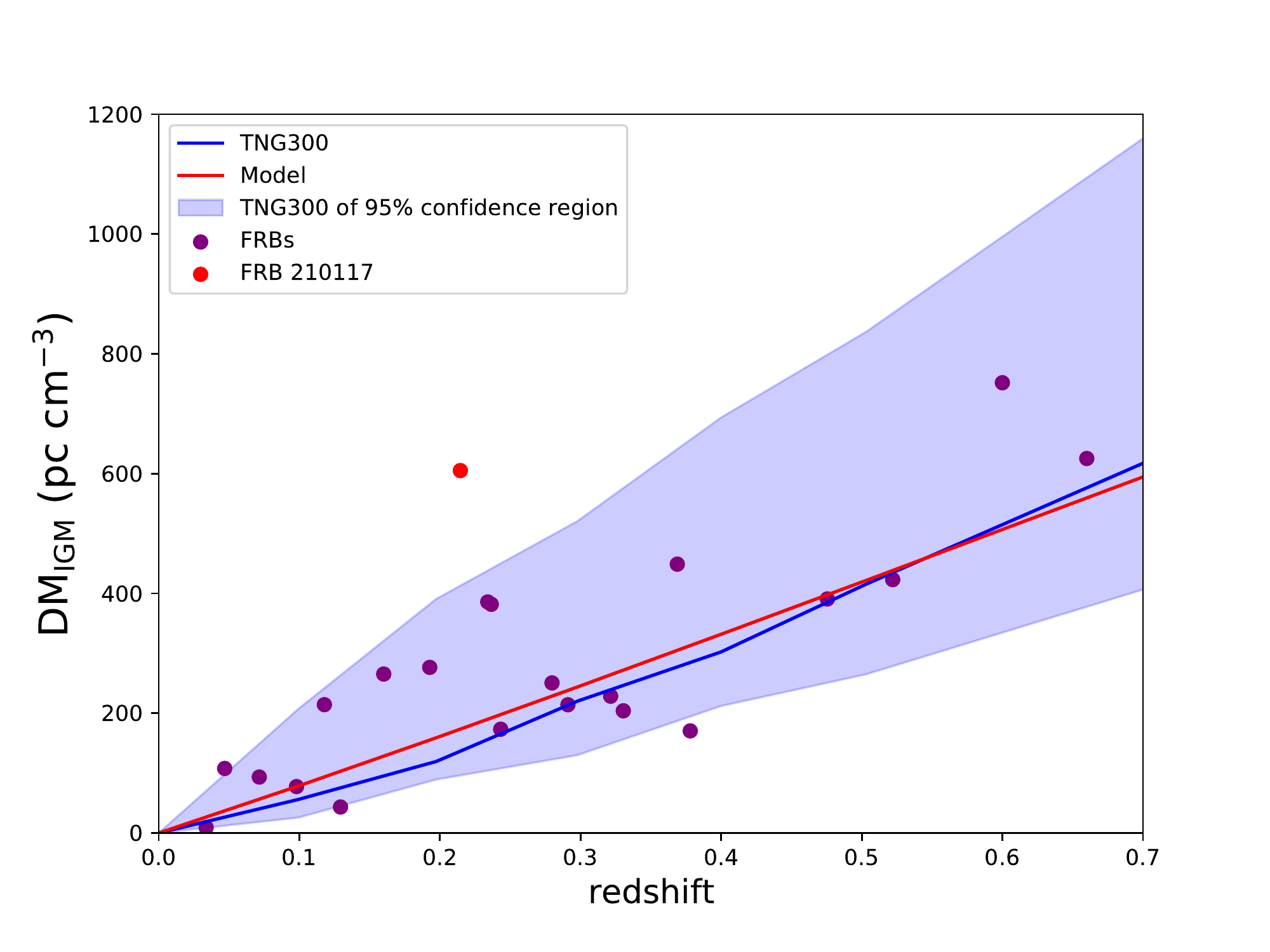}
\caption{\label{DM-z} The ${\rm DM}_{\rm IGM}-z$ relation for FRBs. The scatter star points are 23 localized FRBs. $\rm DM_{IGM}$ values are derived by deducing the $\rm DM_{MW}$ ,$\rm DM_{halo}$ and $\rm DM_{host}$ from $\rm DM_{obs}$. $\rm DM_{MW}$ values are deduced from NE2001 model, and $\rm DM_{halo}$ values are estimated to be $65 \ \rm pc\ cm^{-3}$. We use the median
value of $\rm DM_{host}$ at different redshifts according to equation (\ref{Phost}). The blue line is the result derived from IllustrisTNG 300 cosmological simulation and the green shaded region is the 95\%
confidence level.The red line represents the theoretical model DM$_{\rm IGM}$ (equation (\ref{DM-z})) assuming $f_{\rm IGM}=0.82,\ \Omega_b=0.0488$ and $H_0= 70 ~\rm km~\ s^{-1} ~Mpc^{-1}$. Due to FRB 210117 (red point) deviates from the 95\% confidence region of DM$_{\rm IGM}$, we exclude it in the following analysis.}
\end{figure}

Though over 600 FRBs have been observed, only 24 of them have been localized. 19 of them are collected at \url{http://frbhosts.org/#explore}, including the nearest one FRB 20200120E. This FRB is localized in a globular cluster in M81, whose distance is only 3.6 Mpc \citep{2021ApJ...910L..18B, 2021arXiv210511445K}. It is obvious that it contains little cosmological information and we exclude it. The other five FRBs localized by the Australian Square Kilometre Array Pathfinder (ASKAP) are given in \cite{James22}.
The parameters of all 23 localized FRBs are shown in Table 1. It lists the redshifts, DM$_{\rm obs}$, DM$_{\rm MW,ISM}$, host type and references of them.
In figure \ref{DM-z}, we plot the $\rm DM_{IGM}-z$ relation of 23 localized FRBs. The derived $\rm DM_{IGM}$ of FRBs are shown as scatters. The red line shows the averaged value of $\rm DM_{IGM}$ from the equation (2). The blue line is the $\rm DM_{IGM}$ estimated from the IllustrisTNG simulation with 95\% confidence region \citep{zzj21}. FRB 210117 deviates from the 95\% confidence region, which may be caused by the DM contributed by the plasma near the FRB source, similar as that of FRB 190520B \citep{Zhao2021,Katz2022}. Therefore, we exclude this FRB in the following analysis.
%Besides, the position of the first 18 FRBs on the celestial sphere are plotted using galactic coordinates and shown in Fig \ref{mollweide} (Positions of the five FRBs in \cite{James22} are not given in that article). This is to explain the rationality of setting the upper and lower limits of $\rm DM_{halo}$ to $ 30 \sim 100\rm \  pc \ cm^{-3}$ in equation (\ref{dmhalo}).}
We use the NE2001 model to calculate the DM contribution by the ISM of Milky Way \citep{NE2001}.
%The standard flat $\Lambda$CDM model with $\Omega_m=0.315 \pm {0.007}$ derived from Planck CMB observations is used.
%such as $\Omega_m=0.315 \pm {0.007},\rm H_0=67.36 \pm {0.54} \ \rm km\ s^{-1} Mpc^{-1} $, and $\Omega_{b}h^2=0.02237 \pm {0.00015}$ \citep{2020A&A...641A...6P}. After deducing $DM_{MW}$ and $DM_{halo}$, we can obtain DM$^{\prime}_{FRB}$ of each FRB.
\begin{figure}
\centering
\includegraphics[width=1.0\textwidth]{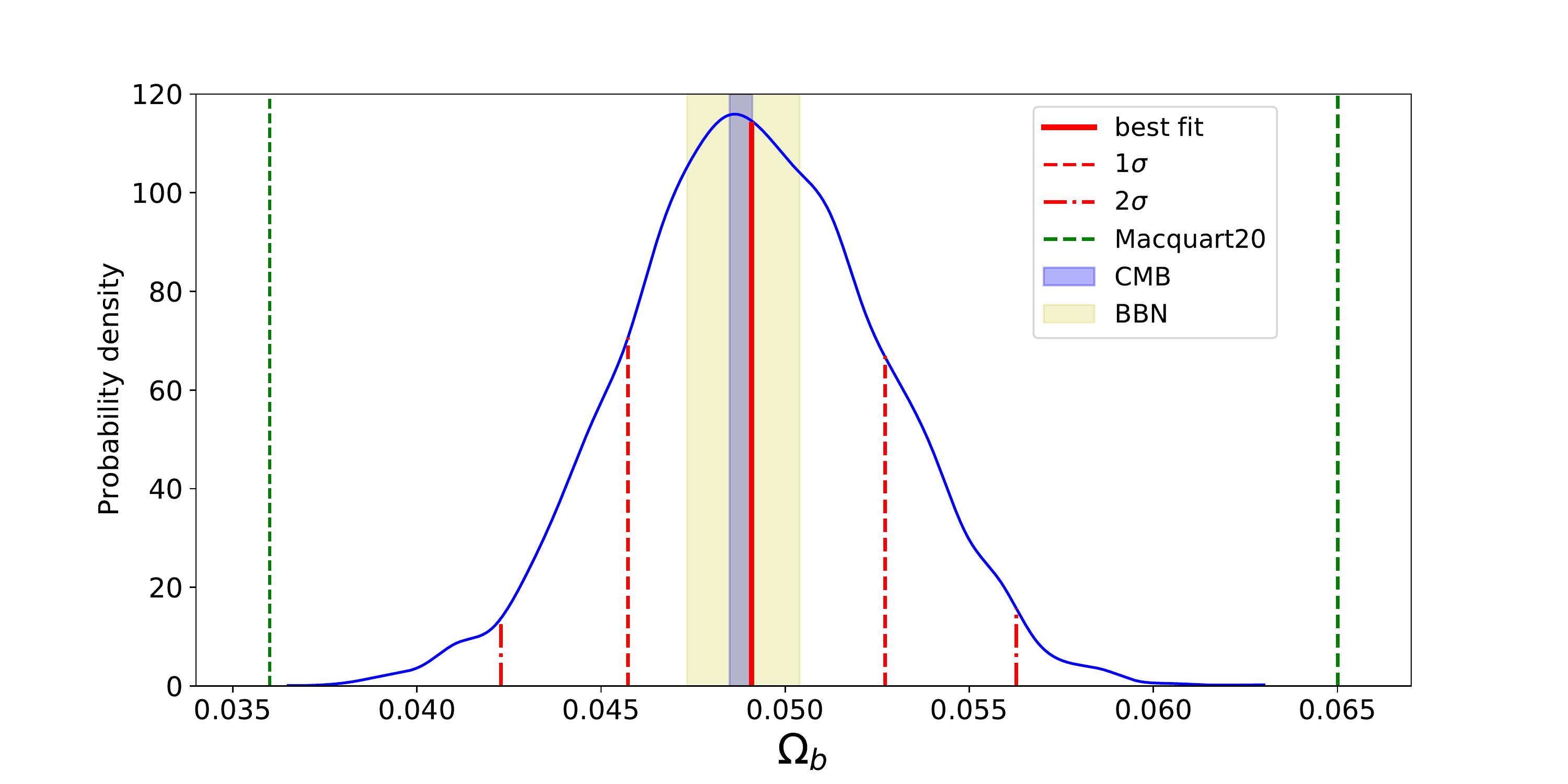}
\caption{\label{PDF}The normalized probability density distribution of $\Omega_b$ from 22 localized FRBs. The best fit is $\Omega_b=0.0490^{+0.0036}_{-0.0033}$ in 1$\sigma$ confidence level.
The red solid line shows the best-fit value, and dotted lines present the 1$\sigma$ and 2$\sigma$ confidence levels. The blue shade region shows the constraint on $\Omega_b$ from CMB with 1$\sigma$ confidence level, and the yellow shade region represents the constraint from BBN with 1$\sigma$ confidence level. The green dashed lines correspond to the 1$\sigma$ uncertainty range of $\Omega_b$ derived by \cite{Macquart20}.}
\end{figure}

\begin{figure}
	\centering
	\includegraphics[width=0.8\textwidth]{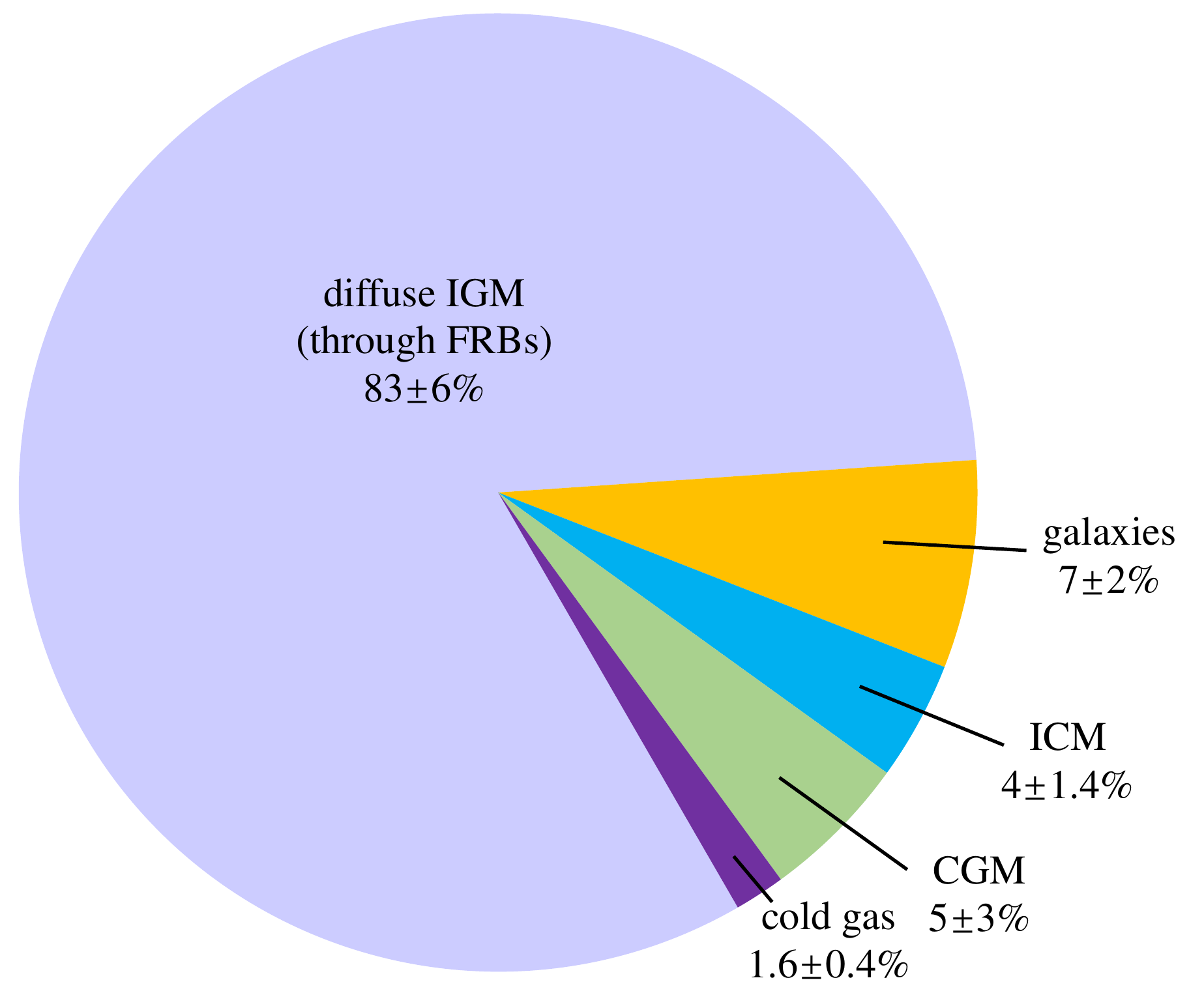}
	\caption{\label{pie}Current observational measurements of the low redshift baryon census. From the DM$-z$ relation of FRBs, 83\% $\pm$ 6\% of the cosmic baryons in the diffuse form are derived.
		The values in collapsed forms, including in galaxies, the CGM, the ICM, and cold gas, are adjusted from \cite{shull12}. Collapsed
		phases (galaxies, CGM, ICM, and cold neutral gas) contribute 17\% $\pm$ 4\% of the total cosmic baryons. From our study, almost 100\% cosmic baryons are found in the late-time Universe.}
\end{figure}

The MCMC Python module \textit{emcee} is used to estimate $\Omega_b$ \citep{Foreman-Mackey2013}. $\Omega_m$ and $H_0$ are also free parameters to be fitted in our analysis. Cosmological observations have tightly constrained them. As shown in below, they have little effect on the final constraints on $\Omega_b$.
The steps of MCMC analysis are as follows.

(a) Firstly, we get the (${\rm DM}_{{\rm FRB}^{\prime}}$,$z$) dataset by deducing the $\rm DM_{MW}$ obtained from the NE2001 model from $\rm DM_{obs}$. Then we calculate the probability density of ${\rm DM}_{{\rm FRB}^{\prime}}$ for each FRB using equations (\ref{dmhalo}), (\ref{PIGM}), (\ref{Phost}) and (\ref{Ptotal}). The joint likelihood function is obtained by multiplying the PDF of each FRB with equation (\ref{likelihood}).

(b) Next, we need to set the priors for the parameters to be fit. For $\Omega_b$, we assume a uniform prior of range [0.03, 0.065]. $H_0$ and $\Omega_b$ are also set priors to estimate their impact. For the Hubble constant $H_0$, the measurements of $H_0$, from SNe Ia and CMB, are mainly distributed between 67 $\rm km\ s^{-1} Mpc^{-1}$ and 73 $\rm km\ s^{-1} Mpc^{-1}$  \citep{2020A&A...641A...6P,Riess2022}. Therefore, we assume a uniform prior of range [67, 73] $\rm km\ s^{-1} Mpc^{-1}$. A fiducial value of $H_0$=70 km/s/Mpc is used by \cite{Macquart20}. Finally, we assume a uniform prior of [0.28, 0.35] for $\Omega_m$ because most of the measurements of $\Omega_m$ are in this range \citep{Pantheonplus,2018ApJ...859..101S}.

(c) At last, we run 10000 steps of MCMC using Python package emcee with the likelihood function equation \ref{likelihood}.
%First, 1000 $\Omega_b$ points are uniformly generated between 0.030 and 0.060. We calculate the probability of $\Omega_b$ with equation (\ref{Ptotal}). The probability density function of $\Omega_b$ is shown in Figure \ref{PDF}.
The result is $\Omega_b=0.0490^{+0.0036}_{-0.0033}$ (1$\sigma$) with a 7.0\% precision. This value is consistent with previous results in 1$\sigma$ confidence level, such as CMB and BBN \citep{2020A&A...641A...6P,Cooke2018}. \cite{Macquart20} used five FRBs to constrain $\Omega_b$, and found $\Omega_b=0.051^{+0.021}_{-0.025}$ with 95 percent confidence by assuming $H_0$=70 km/s/Mpc. This value is compatible with our result in 1$\sigma$ confidence level, but with large error.

In the calculation, the fraction of baryons in the IGM $f_{\rm IGM}=0.82$ is used. It must be noted that $\Omega_b$ and $f_{\rm IGM}$ are degenerated. In fact, only the product of $\Omega_b$, $H_0$ and $f_{IGM}$ can be derived. Conversely, if the value of $\Omega_b$ from CMB is fixed, the constraint on $f_{\rm IGM}$ should be close to 0.82. The same prior for $H_0$ and the best constraint on $\Omega_b$  from CMB are used. Using the method as above, we derive $f_{\rm IGM}=0.83\pm 0.06$.
%In their analysis, 6 localized FRB are used, and there are three parameters in addition to $\Omega_b$ to be fitted: $F$, the parameter quantifying the strength of the baryon feedback, $\mu$ and $\sigma_{host}$, while these parameters are obtained from a model based on the simulation of \cite{zgq20} in our work. More parameters to be fitted and less data may increase the error. The cosmic microwave background and Big Bang nucleosynthesis give tight constraints on $\Omega_b$. Limits on $\Omega_b$ from FRBs are independent of but less accurate than these measures so far. Implementing the restriction with our approach may require more FRBs.

Figure \ref{pie} shows the current census of baryons
in the low-redshift Universe, including diffuse form revealed by FRBs, and collapsed forms. Collapsed forms contain galaxies, the circumgalactic medium (CGM), intercluster medium
(ICM), and cold neutral gas (H I and He I). These slices show the
contributions, $\Omega^i_b/\Omega_{\rm tot,b}$ to the total baryon content from components $i$. Previous studies found that $5\% \pm 3\%$ may reside in circumgalactic gas, 7\%$\pm$2\% in galaxies,
1.7\%$\pm$0.4\% in cold gas and 4\%$\pm$1.4\% in clusters \citep{shull12}. From our analysis, the measurement of FRBs can account for 83\% $\pm$ 6\% of the baryons in the Universe. Therefore, the missing baryons are found in the IGM through localized FRBs.

\section{Discussion and Conclusions}

The statistical and systematic errors must be taken into consideration. The statistical error refers to the small number of localized FRBs. As mentioned above, this will be improved with the increase of localized FRBs.
%We also show that 100 mock FRBs can give a tighter limit.
The systematic error includes all the uncertainties except statistical errors, such as cosmological parameters, observational errors of DM and redshift. It cannot be eliminated by more data. More precise measurements of these parameters are needed. Besides, YMW16 model gives a larger DM$_{\rm MW}$ value than the NE2001 model. This may lead to a similar but smaller result.
%We also repeat our analysis with the
%$\rm DM_{MW}$ estimated by the YMW16 model, and obtain a similar result, i.e., $\Omega_b=*****\pm *****$, which is smaller than the above one.

In equation (\ref{PIGM}), the best-fit parameters for $C_0$, $\sigma_{DM}$ and $A$ are taken from \cite{zzj21}. Their uncertainties may affect the final $\Omega_b$ constraint. We derive their uncertainties from the original data. From equations (\ref{PIGM}) (\ref{Ptotal}) and (\ref{likelihood}), it is obvious that the uncertainty of $A$ has no effect on the result. To consider the effect of $C_0$ and $\sigma_{DM}$, we derive $\Omega_b$ using the best-fit values of $C_0$ and $\sigma_{DM}$ plus or minus the uncertainties respectively, totaling four probability distributions of DM$_{\rm IGM}$. The results of fits are $\Omega_b=0.0492^{+0.0035}_{-0.0034}$, $0.0492^{+0.0035}_{-0.0034}$, $0.0481^{+0.0036}_{-0.0034}$ and $0.0493^{+0.0036}_{-0.0033}$ in these four case. These constraints are very close to the original result (approximately 0.25$\sigma$) and their errors are almost the same. Therefore we can conclude that uncertainties of $C_0$, $\sigma_{DM}$ and $A$ have almost no effect on final $\Omega_b$ constraint.
Besides, our calculation is not entirely independent of cosmological parameters, such as $\Omega_m$ and $H_0$. We adopt their ranges limited by observations.
%The cosmic matter density $\Omega_m=0.315 \pm {0.007}$ from Planck CMB observation is used.
%A similar value is derived from type Ia supernovae \citep{Scolnic2018}. The precise $\Omega_m$ (relative error $\sim 2\%$) has a little effect on the derived $\Omega_b$. CMB occurred in the early era of our universe whereas the observed FRBs are at low redshifts. Cosmological parameters obtained from cosmic observations at different times may vary. For instance, $H_0$ derived by CMB observations \citep{2020A&A...641A...6P} differs from that measured by local distance ladders \citep{Riess2022}, which is the so-called ``Hubble Tension". In the calculation, a fiducial value of $H_0=70$ km/s/Mpc is used. From observations, the error of $H_0$ is about 1\% \citep{2020A&A...641A...6P,Riess2022}, which has a little effect on our results. %However, $\Omega_b$ is not measured at the low-z universe. It means that we have to constrain a parameter of the low-z universe with the part of high-z data. We can pin our hope on another novel method of giving $\Omega_b$ at the low redshift.

In summary, we estimate the cosmic baryon density, $\Omega_b$ using 22 localized FRBs by considering the probability distributions of DMs contributed by host galaxies and IGM.
We obtain the result $\Omega_b=0.0492^{+0.0035}_{-0.0034}$ (1$\sigma$) with a 7.0\% precision. The result is consistent with those derived from CMB and BBN. So, FRBs can provide an independent measurement of $\Omega_b$ with high precision. Our work shows that the missing baryons are found in diffuse IGM by FRBs.

\section*{acknowledgements}
We thank the anonymous referee for helpful comments. This work was supported by the National Natural Science
Foundation of China (grant Nos. U1831207 and 12273009), the National Key Research and Development Program of China
(2022SKA0130100), the Fundamental Research Funds for the Central Universities (No. 0201-14380045) and the China Manned Spaced Project (CMS-CSST-2021-A12).

\bibliography{sample}

\begin{table*}
\vspace{-1.5cm}
\centering
\caption{Properties of localized FRBs}
\begin{tabular}{cccccc}
\hline
Name & Redshift & DM$_{\rm obs}$ &DM$_{\rm MW,ISM}$ & Host Type & Reference \\
 &  & $(\rm pc \ cm^{-3})$ &$(\rm pc \ cm^{-3})$& &  \\
\hline
FRB 121102 &    0.19273 &557	&188.0  & I   & \cite{2016Natur.531..202S}\\
FRB 180301 &    0.3304  &534    &152    & I   & \cite{2021arXiv210801282B}\\
FRB 180916 &	0.0337  &348.8	&200.0  & II     & \cite{2020Natur.577..190M}\\
FRB 180924 &    0.3214  &361.42 &40.5   & III     & \cite{2019Sci...365..565B}\\
FRB 181030 &    0.0039  &103.5  &41     & I     & \cite{2021ApJ...919L..24B}\\
FRB 181112 &    0.4755	& 89.27 &102.0  & III     & \cite{2019Sci...366..231P}\\
FRB 190102 &    0.291 	&363.6	&57.3   & III     & \cite{2020ApJ...895L..37B}	\\
FRB 190523 &    0.66    &760.8  &37.0   & III    & \cite{2019Natur.572..352R}\\
FRB 190608 &    0.1178  &338.7	&37.2   & III     & \cite{2021ApJ...922..173C}	\\
FRB 190611 & 	0.378   &321.4	&57.8   & III     & \cite{2020MNRAS.497.3335D}  \\
FRB 190614 &    0.6     &959.2  &83.5   & III       & \cite{2020ApJ...899..161L}\\
FRB 190711 &    0.522 	&593.1  &56.4   & I     & \cite{2020ApJ...903..152H}\\
FRB 190714 &    0.2365	& 504.13&38.0   & III     & \cite{2020ApJ...903..152H}\\
FRB 191001 &	0.234   &507.9  &44.7   & III     & \cite{2020ApJ...903..152H}\\
FRB 191228 &    0.2432  &297.5  &33     & III     & \cite{2021arXiv210801282B}\\
FRB 200430 &    0.16	&380.25 &27.0   & III     & \cite{2020ApJ...903..152H}\\
FRB 200906 &    0.3688  &577.8  &36     & III     & \cite{2021arXiv210801282B}\\
FRB 201124 & 	0.098   &413.52 &123.2  & II     & \cite{2021ApJ...919L..23F}\\
FRB 210117 &    0.2145  &730    &34.4   & III     & \cite{James22}\\
FRB 210320 &    0.2797  &384.8  &42    & III     & \cite{James22}\\
FRB 210807 &    0.12927 &251.9  &121.2  & III     & \cite{James22}\\
FRB 211127 &    0.0469  &234.83 &42.5   & III     & \cite{James22}\\
FRB 211212 &    0.0715  &206    &27.1   & III     & \cite{James22}\\

\hline
% \toprule
% \toprule
 \end{tabular}
 \label{data}
 \vspace{0.5cm}
\end{table*}

\end{document}